\documentstyle[aps,prl,multicol,psfig]{revtex}
\tighten
\begin{document}
%\draft
\title{Spectral properties of incommensurate charge-density wave systems}
\author{G. Seibold$^{*,\dagger}$, F. Becca$^{*,\sharp}$, F. Bucci$^*$, 
	C. Castellani$^*$, C. Di Castro$^*$, and M. Grilli$^*$}
\address{$^*$ Istituto Nazionale di Fisica della Materia e
Dipartimento di Fisica, Universit\`a di Roma ``La Sapienza'',\\
Piazzale A. Moro 2, 00185 Roma, Italy}
\address{$^{\dagger}$ Institut f\"ur Physik, BTU Cottbus, PBox 101344, 
         03013 Cottbus, Germany}
\address{$^{\sharp}$ Istituto Nazionale Di Fisica Della Materia and
International School for Advanced Studies,\\ 
Via Beirut 4, 34013 Trieste, Italy}

\date{\today}

\maketitle

\begin{abstract}
The concept of frustrated phase separation is applied to
investigate its consequences for the electronic structure
of the high T$_{c}$ cuprates. The resulting incommensurate
charge density wave (CDW) scattering is most effective in creating
local gaps in k-space when the scattering vector connects states
with equal energy. Starting from an open Fermi surface we find that
the resulting CDW is oriented along the (10)- and (or) (01)-direction
which allows for a purely one-dimensional or a two-dimensional
``eggbox type'' charge modulation. In both cases the van Hove
singularities are substantially enhanced, and 
the spectral weight of Fermi surface states near the M-points,
tends to be suppressed. Remarkably, a leading edge gap arises
near these points, which, in the eggbox case, leaves finite
arcs of the Fermi surface gapless. 
We discuss our results with repect to possible consequences for 
photoemission experiments.
\end{abstract}

\pacs{}

\begin{multicols}{2}
\section{Introduction}
Striped phases are an important issue in the
discussion on physical properties of the high-T$_c$ materials
\cite{CAST,EMERY1}. 
Incommensurate magnetic peaks displaced 
from the antiferromagnetic wave-vector along $\bf k_x$ and $\bf k_y$ 
have been observed by inelastic neutron scattering in both 
La$_{2-x}$Sr$_x$CuO$_4$ \cite{CHEONG,MASON,THURST} and 
YBa$_2$Cu$_3$O$_{7-x}$ (YBCO) \cite{MOOK1}, whereas in the 
Bi$_2$Sr$_2$CaCu$_2$O$_8$
material the incommensurate magnetic peak position has not yet been
resolved \cite{MOOK2}. However, incommensurate magnetic scattering
may result from either some kind of spiral phase (involving a modulation 
of the transversal spin components only) or from longitudinal spin 
fluctuations where in this case a strong coupling to
the charge is expected. Evidence for cooperative charge- and spin scattering
is based on the observation of domain walls in
La$_{1.6-x}$Nd$_{0.4}$Sr$_{x}$CuO$_{4}$ \cite{TRAN} where 
the low temperature tetragonal lattice
structure and the filling close to $1/8$ are suited to pin
the density fluctuations, giving rise to a static CDW phase
as revealed by commensurate ionic shifts observed in neutron scattering.
From the temperature dependence of
the charge- and spin-order peaks one can further conclude
that the stripe order is driven by the charge rather than
by the spins. 
Additional support for charge induced stripe correlations is provided
by recent neutron scattering studies of the phonon dispersion in
YBCO \cite{MOOK3}. These measurements show a large broadening
in the spectrum at wave vectors which are twice the magnetic
incommensurability. Moreover, the broadening in the phonon spectrum
occurs at higher temperatures as the incommensurable magnetic peaks
indicating again that charge fluctuations are the `driving force'
behind  incommensurable magnetic scattering

The formation of stripe correlations in these compounds can be well
understood within the concept of frustrated phase
separation \cite{EMERY2} where a phase separation  instability
is prevented by long-range Coulomb interactions \cite{notaps}. As a result 
the long-wavelength density fluctuations associated with 
phase separation are
suppressed in favor of shorter-wavelength density fluctuations,
giving rise either to dynamical slow density modes \cite{EMERY2}
or to incommensurate charge density waves (CDW) \cite{CAST,BECCA}.
Concerning the pairing mechanism the theoretical 
proposals differ with regard to the role
of the stripe correlations on the superconducting pair formation. 
Whereas some theories consider the formation of quasi 1-d `rivers of charge'
as a necessary environment for the quasiparticles to pair or to
enhance T$_c$ \cite{EMERY1,BIANC}
others take the fluctuations associated with the 
stripe instability itself as a pairing mechanism \cite{CAST}.

The proximity to a stripe instability tuned by doping
and temperature allows to interpret many properties of the 
cuprates as due to a Quantum Critical Point (QCP)
\cite{notaQCP} located near optimal doping
\cite{CAST,PERALI,CDGZP,CAPECOD}. Within this scenario
the singular scattering induced by the critical
fluctuations would be responsible for both the anomalous
normal-state properties and the large superconducting
critical temperatures. Then the phase diagram of the cuprates
is partitioned in the (nearly) ordered, the quantum critical, and 
the quantum disordered regions corresponding to
the under-, optimally, and over-doped regions of the phase
diagram of the cuprates. 

The relevance of the stripe phase and 
of the stripe fluctuations should find an experimental
confirmation (or a disproof) in the analysis of the 
single-particle spectral properties.
Taking the point of view that incommensurate CDW scattering
contributes to determine
the electronic properties of the high-T$_c$ materials 
the question arises how the electronic structure around the Fermi
energy is affected by this scattering and if this can be detected by 
angle-resolved photoemission spectroscopy (ARPES) \cite{shendessau}.
In fact, there are several features which seem to be
common to all p-type cuprates 
(see e.g. Ref.\ \cite{RANDERIA} and references therein). 
In particular, the dispersion displays
an extended saddle-point at ($\pi/a,0$)  (here and in the following
we will consider unitary lattice spacing $a=1$)
with an energy close to E$_{F}$, which
has been observed in Bi2212 \cite{DESSAU} as well as in
YBaCuO \cite{GOFRON} samples.
Furthermore, the Fermi surface 
of Bi2212 is very strongly nested with a nesting vector of
approximately 0.9 ($\pi/a,\pi/a$). The resulting picture
for the Fermi surface therefore consists of tube-like
structures around the M-points which are connected by more
or less straight pieces.  
It has been shown in Ref.\ \cite{SALKOLA} that many of these unusual 
Fermi surface features can be explained by assuming a disordered stripe
phase to be present in the high-T$_c$ copper oxides. The fact 
that the charge carriers are exposed to a stripe potential mirrors
in the Fermi surface through one-dimensional characteristics which
predominantly appear around the M-points.

A further fascinating Fermi surface feature of the high T$_c$ superconductors,
which has been detected in ARPES measurements concerns the pseudogap
phenomenon which is observed in optimally and underdoped
bismuth compounds below a doping dependent temperature $T^*>T_c$
\cite{MARSHALL2,DING,HARRIS,BIANC3}. 
In underdoped samples of Bi2212 \cite{MARSHALL2,DING}
there is a qualitative change in
the electronic structure with respect to the optimally doped
system already above T$_{c}$, with the appearance of
so-called leading-edge gaps around the M points.
In these k-space regions there is no indication of a quasiparticle peak,
and the spectra are dramatically broadened
and shifted to higher binding energies with a maximum at
100-200 meV. As a consequence
large portions of the Fermi surface around
these points are not visible whereas along the $\Gamma \rightarrow X$
directions the spectral lineshapes are similar to those
of the optimally doped samples.  
Similar results where obtained in the one-plane
material Bi2201 \cite{HARRIS} showing that bilayers are not
essential to the pseudogap phenomena.

A common explanation for the features described above is based on
the existence of pre-formed pairs already above T$_{c}$, since 
a continuous evolution of the spectra from the normal
to the superconducting state \cite{MOHIT} was apparently observed.
In particular, a superconducting d-wave gap 
is in accordance with the symmetry of the pseudogap. 
Moreover, for materials with $k_F \xi_0 \lesssim 10$ the existence of two 
temperature scales for pairing and phase coherence is a 
rather natural consequence. However, this simple scenario is
put in jeopardy by various experimental findings. First of all
the rapid increase of $T^*$ by decreasing doping
is to be contrasted with the much more moderate increase
of the maximum gap at low temperature.
($T^*$ doubles when $T_c$ goes from 90 K to about 75 K in Bi-2212,
whereas in the same doping range, $\Delta(T=0)$ around the M-points
changes by thirty per cent at most). This suggests that a substantial
temperature dependence of the pairing potential is present in this
doping range. Furthermore recent ARPES experiments \cite{NORMAN} show that 
the pseudogap above $T_c$ gradually appears at the M points leaving
extended segments of the Fermi surface gapless. These segments
continuously shrink to a point-like node only below the
superconducting critical teperature $T_c$. Needless to say
that these behaviors do not correspond  neither to the
usual BCS gap opening process nor to the phase locking of uniform
preformed pairs. Instead it might find an interpretation
along the line of coexisting pairs and fermionic quasiparticles
\cite{RANNINGER,GESHKENBEIN}.

In front of this unsettled issue, 
one may ask whether the rich and unexplained behavior of the 
pseudogap phenomena could result from an
electron-hole pair scattering above T$_{c}$ rather than from 
pair correlations alone. This line of argumentation has been followed
in several analyses based on the scattering between fermions
and collective spin excitations
\cite{pines,chubukov,SHEN}. A quite similar outcome 
would instead arise from the quasiparticles
being scattered by incommensurate charge fluctuations \cite{CAPRARA}.
The fact that a substantial part of the
spectral features of the optimally and underdoped
cuprates can be explained in terms of particle-hole scattering
is also supported by recent experiments \cite{BIANC3} measuring
the Fermi surface by sequential angle-scanning photoemission.
These results clearly show the missing segments of the Fermi surface
around the M-points with a shape, which can be interpreted in terms of
scattering of the quasiparticles with quasicritical mixed spin and
charge fluctuations \cite{CAPRARA}.

The idea that the pseudogap could (not only) arise from pairing in the
particle-particle channel, but also from different scattering
mechanims (like CDW fluctuations)
 finds an additional support in the recent analysis of Ref.
\cite{PANAGOPOULOS}, where the experimental results on
the low-temperature penetration depth suggest that 
the doping behavior of the low-temperature gap near the nodes in the
(1,1) direction is similar to what expected from 
standard d-wave BCS theory
with the gap scaling with $T_c$. This is in contrast with 
the behavior of the pseudogap measured at
the k-points around $(\pm \pi/a, 0)$ and $(0, \pm \pi/a)$ 
in ARPES experiments
below $T^*$, which increases with decreasing $T_c$. 
This suggests the possibility that at high temperature
($T^*>T>T_c$) the gap is due to CDW fluctuations scattering
in the particle-hole channel.
In this regard it was shown in Refs.\ 
\cite{CAST,BECCA} that the incommensurate CDW 
scattering resulting from the competition between 
phase separation and long range repulsive Coulomb forces, 
occurs near some scattering 
vector ${\bf q}_c$ not related to the Fermi vector k$_F$. 
It was also shown
within a standard large-N approach \cite{CAST,BECCA} that
the effective scattering amplitude close to ${\bf q}_c$ in the
particle-hole channel is  
strongly attractive and similarly structured in momentum space 
in the particle-particle channel also, thus resulting 
in substantial pairing in the d-wave channel. 
Within this scenario, the strong  momentum-dependent effective interaction
is responsible both for (preformed) pairing and for pseudogap formation
due to dynamical charge modulation. Therefore 
searching for spectral features arising from scattering mechanisms
acting in the particle-hole channel is not necessarily in contrast
with the interpretation of other features in terms of pairing
in the particle-particle channel. 

In the present paper we are interested in the question if 
incommensurate CDW scattering,
resulting from the competition between phase separation and long range
repulsive Coulomb forces, can account for 
 some of the above anomalous features observed in
photoemission experiments. 
This problem was partly tackled in Ref.\cite{CAPRARA} within a perturbative
treatment of quasicritical spin and charge fluctuations. This perturbative
approach is unfortunately not feasible in the underdoped region,
deep inside the stripe phase. Therefore we perform here a mean-field
analysis, which should capture the non-perturbative character
of well-formed stripes. Of course, the dynamical character of the stripe
fluctuations will be missed. Nevertheless, at short distances and time
as those probed in photoemission experiments, this description should be
appropriate. 

We emphasize once again that the complexity of the physics of the cuprates,
particularly of the underdoped ones, involves spin and Cooper as well as
charge density fluctuations. Our aim is therefore {\it not}
to account for all the spectral properties of these materials,
which likely arise from the interplay of all these mechanisms.
We rather conservatively aim to point out that some of the 
prominent (and puzzling) features of the cuprates could find
a natural interpretation by assuming charge quasi-ordering.

The paper is structured as follows. In Section II the model and the general
formalism are presented. Section III contains the physical results
for the Fermi surface and the spectral densities, which
are discussed and summarized in the conclusive Section IV.

\section{Formalism}
Starting point is the Hubbard-Holstein model with an
additional in-plane long-range Coulomb interaction
\end{multicols}
\begin{eqnarray}
H&=&-t\sum_{<i,j>,\sigma}(c_{i\sigma}^{\dagger}c_{j\sigma} + H.c.)
-t'\sum_{<<i,j>>,\sigma}(c_{i\sigma}^{\dagger}c_{j\sigma} + H.c.)
+g\sum_{i\sigma}(A_{i}^{\dagger}
+A_{i})(n_{i\sigma}-<n_{i\sigma}>) \nonumber \\
&+&\omega_0\sum_{i}A_{i}^{\dagger}A_{i}-\mu_0 \sum_{i\sigma}n_{i\sigma}
+U\sum_{i}n_{i\uparrow}n_{i\downarrow}+\frac{1}{2N}\sum_{q}\frac{V_{c}}
{\sqrt{G^2(q)-1}}\rho_{q}\rho_{-q} \label{HUHOL}
\end{eqnarray}
\begin{multicols}{2}
where  $c_{i\sigma} (c_{i\sigma}^{\dagger})$ destroys
(creates) an electron with spin $\sigma$ at site i and $A_i (A_i^{\dagger})$
destroys (creates) a local Holstein-type phonon at site i. 
The summation over nearest- and next-nearest neighbour
sites is indicated by $<i,j>$ and $<<i,j>>$, respectively.
$\sum_{\sigma}n_{i\sigma}=\sum_{\sigma}c^{\dagger}_{i\sigma}
c_{i\sigma}$ is the local electron density and its Fourier transform
is given by $\rho_{q}=\sum_{k\sigma}c_{k+q\sigma}^{\dagger}c_{k\sigma}$.
The last term in Eq.(\ref{HUHOL}) describes the Coulomb interaction 
between electrons on a two-dimensional, square lattice 
(lattice constant $a$ in x- and y-direction), which is considered
as a plane embedded in the 3-dimensional lattice (plane distance
$d$ in the z-direction).
The dielectric constants in the plane and perpendicular to it
are given by $\epsilon_{\|}$ and $\epsilon_{\perp}$, respectively. 
The Coulombic coupling constant is $V_{c}=\frac{e^2d}{2\epsilon_{\perp}a^2}$. 
On the z=0 plane the momentum dependence of the Coulomb
potential is found to be
$G(q_{x},q_{y})=\lbrack \epsilon_{\|}/\epsilon_{\perp}(a/d)^2\rbrack
\lbrack \cos(aq_{x})+\cos(aq_{y})-2 \rbrack-1$. 
As usual, the sum in the Coulombic potential does not include the
zero-momentum component, since we are supposing that the diverging
$q=0$ interaction between electrons is canceled by the contribution of
a uniform positively charged ionic background.

Since we are interested in the limit of strong local repulsion,
we take the limit $U \rightarrow \infty$, which gives rise
to the local constraint of zero double occupancy
$\sum_{\sigma}n_{i\sigma}\le 1$.
To implement this constraint we use a standard slave-boson technique
\cite{SLABOS,RADGAUGE} by performing the usual substitution  
$c_{i\sigma}^{\dagger}
\rightarrow c_{i\sigma}^{\dagger}b_{i}$, $c_{i\sigma}
\rightarrow c_{i\sigma}b_{i}^{\dagger}$ and implement the
local constraint by a local Lagrange multiplier 
$\lambda_{i}$.
The model can first be solved in the mean field
approximation by setting the $b_{i}^{(\dagger)}$ and $\lambda_{i}$ 
bosons to their constant self-consistent values $b_0$ and $\lambda_0$,
respectively. At this stage the phonons and the Coulombic interaction
decouple from the fermionic quasiparticles for an homogeneous
mean-field solution. 
The system then results in free quasiparticles with a shifted
chemical potential $\mu=\mu_{0}-\lambda_{0}$ and a dispersion
$E_{k}=-2tb_{0}^2\epsilon_{k}$, where
$\epsilon_{k}=\lbrack \cos(ak_{x})+\cos(ak_{y}\rbrack
+t'/t\lbrack \cos(ak_{x}+ak_{y})+\cos(ak_{x}-ak_{y}) \rbrack$
and $b_{0}^2=\delta$ corresponds to the concentration
of doped holes.
The mean-field value for $\lambda_{0}$ is determined by 
$\lambda_{0}\equiv \lambda_{0}^{0}+(t'/t) \lambda_{0}^{1}
=2t\sum_{k}f(E_{k})\epsilon_{k}$ where $f(E)$ is the
Fermi function. 

The effective interaction leading to scattering between
quasiparticles arises from the exchange of the bosonic
fields beyond the mean-field approximation.
Working within the radial gauge \cite{RADGAUGE} 
one can define a three-component 
field $\alpha^{\mu}=(\delta r, \delta \lambda, \delta \phi)$ where
$\delta\phi$ is the lattice displacement field, and 
$\delta r$, $\delta \lambda$ are the fluctuating parts of the modulus
of the $b_{i}^{(\dagger)}$-bosons and of the Lagrange
multiplier $\lambda_{i}$ respectively.
The static bare scattering amplitude in the particle-hole
channel can be written as \cite{BECCA}
\begin{equation}\label{effscatt}
\Gamma_0(k,k';q)=-\sum_{\mu\nu} \Lambda^{\mu}(k',-q)D_0^{\mu\nu}(q)
\Lambda^{\nu}(k,q)
\end{equation}
where the vertices $\Lambda^{\mu}$ coupling the fermionic quasiparticles
to the bosons are defined as
\begin{eqnarray}
\Lambda_{r}(k,q)&=&-2tb_{0}^2(\epsilon_{k+q/2}+\epsilon_{k-q/2}) \nonumber \\
\Lambda_{\lambda}(k,q)&=&i \nonumber \\
\Lambda_{\phi}(k,q)&=& -2g \nonumber 
\end{eqnarray}
and $D_0^{\mu\nu}(q)$ denotes
the matrix of the bosonic Greens function \cite{BECCA}
with all elements being zero except for
\begin{eqnarray} 
D_0^{r,r}& = & (b_0^2/2)\left[\lambda_0^0\lbrack
\sin^2(q_{x}/2)+\sin^2(q_{y}/2) \rbrack \right. \nonumber \\ 
& + &   (t'/t)\lambda_0^1\lbrack \sin^2((q_{x}-q_{y})/2)+
\sin^2((q_{x}+q_{y})/2) \rbrack\left. \right], \nonumber \\ 
D_0^{r,\lambda} & = & D_0^{\lambda,r}=ib_0^2/2, \nonumber \\
D_0^{\phi,\phi} & = & \omega_0.
\end{eqnarray}

We finally obtain a Hamiltonian, describing the effective 
interaction between quasiparticles
\begin{equation} \label{H0}
H=\sum_{k\sigma} E_{k}n_{k\sigma}+\frac{1}{2N}\sum_q V_{q} \rho_{q}\rho_{-q}
\end{equation}
where the static effective interaction is given by
\begin{equation}\label{POT}
V_{q}=\Gamma(k_{F},k'_{F};q,\omega=0) 
     = \tilde{U} + \gamma_{q} -\frac{2g^2}{\omega_{0}}+ \frac{V_{c}}
{\sqrt{G^2(q)-1}}
\end{equation}
$\tilde{U}=-4E_{F}/\delta$ 
 and $\gamma_{q}=\frac{\lambda_{0}^{0}}{2\delta}\lbrack
\sin^2(q_{x}a/2)+\sin^2(q_{y}a/2) \rbrack +
\frac{\lambda_{0}^{1}}{2\delta}\lbrack \sin^2((q_{x}-q_{y})a/2)+
\sin^2((q_{x}+q_{y})a/2) \rbrack$. Within a Fermi-liquid scheme,
$V_q$ represents the effective residual interaction between
the quasiparticles
once the screening due to intraband particle-hole quasiparticle
bubbles is disregarded. 

The Hamiltonian (\ref{H0}) is the starting point of all
further investigations. Its interaction consists of a part,
$\tilde{U}+\gamma_q$,
increasing with $|\bf{q}|$, which describes the residual scattering
of the quasiparticles through the slave-bosons exchange. The superposition
of the long-range Coulombic part causes the interaction to
exhibit a  minimum as a function of ${\bf q}$. The electron-phonon
coupling $2g^2/\omega_0$ rigidly
shifts the potential to lower values. This will induce an instability
in the density-density response function at a critical value of
${\bf q}={\bf q}_{c}$ when $1+V_{{\bf q}_c}
\Pi({\bf q}_c)=0$, where $\Pi({\bf q})$ is the particle-hole
fermionic bubble. This instability marks therefore a transition to an
incommensurate CDW with scattering vector
$\bf q_{c}$ \cite{notainstab}.

To proceed we will approximate the effective model Eq. (\ref{H0})
by a Hartree factorization and assume that the symmetry of the system
is broken with respect to a given vector $\bf q_{c}$ and its multiples.
Notice that retardation effects
arising from the frequency dependence of the total scattering amplitude
 have been neglected at this level. This mean-field treatment
of Eq.\ref{H0} finds some 
justification in investigating static or low-energy
properties of the quasiparticles close to the Fermi level,
whereas this approximation is crude in the analysis of dynamical
properties at frequencies, which are comparable with
the typical energies of the exchanged boson.

In order to reproduce the 'unperturbed' Fermi surface 
(i.e. of the overdoped system) of the 
Bi2212 compounds at optimal doping we take the hopping parameter $t'$
to be $t'=-0.45t$, which leads to an open Fermi surface 
centered around $(\pm \pi/a, \pm \pi/a)$. It turns out that
within a linear response approach as in Ref. \onlinecite{BECCA},
for this kind of bandstructure the CDW instability occurs first
in the (1,0) or (and) (0,1) directions.
In fact, a negative value of $t'$ enhances the density of states 
around the M-points, which mirrors in an enhancement
of the fermionic bubble $\Pi({\bf q})$ 
in the (1,0) and (0,1) directions.
Thus in a RPA-like treatment the divergence of the density-density
response function preferably takes place for
$\bf q = \bf q_{c}$ in the $ (1,0)$ or 
in the $(0,1)$ directions. 
Within a linear-response analysis, the instability is forced
to occur at a given wavevector and the above arguments are
in favor of the instability taking place  with a modulation 
alternatively in the $x$ or $y$ direction described by the
order parameter  $<\rho_{q}>^{1D}=
\sum_{n}<\rho_{q}>\delta_{q,nq_{c}^{x/y}}$.
However, a non-linear instability
(as the one here described by a standard set of Hartree 
mean-field equations) can also occur simultaneously
along both directions leading to
a two-dimensional eggbox type modulation of the charge.
In this case, the density expectation value is given by
 $<\rho_{q}>^{egg}=\sum_{n}<\rho_{q}>\lbrack \delta_{q,nq_{c}^{x}}
 + \delta_{q,nq_{c}^{y}}\rbrack$.

Although we are now dealing with an effective one-particle
Hamiltonian, it is clear that the system cannot be diagonalized
for general incommensurate $\bf q_{c}$ vectors. 
We therefore represent $\bf q_{c}$ as
\begin{equation}\label{repqc}
\bf q_{c}= \frac{\pi}{a} (\frac{n_{x}}{m_{x}};\frac{n_{y}}{m_{y}})
\end{equation}
which allows us to simulate the 'incommensurability' by increasing
the values for both $n_{i}$ and $m_{i}$, respectively.
Once the the scattering term in Eq. \ref{H0} is decoupled 
{\it \`a la} Hartree, for each k-point in the reduced BZ, the bare 
Bloch functions $\Psi_{\bf k}(n)\equiv \Psi({\bf k}+n\bf{\bf q}_c)$ 
with k-vector ${\bf k},{\bf k}+{\bf q}_c,{\bf k}+2{\bf q}_c,....,
{\bf k}+n_{max}{\bf q}_c$, are mixed by the interaction. 
$n_{max}$ depends on ${\bf q}_c$
via the condition ${\bf k}+(n_{max}+1){\bf q}_c={\bf k}$.
For each k-point, the Hamiltonian can be put in a matrix
form, which can be diagonalized thus finding the linear
transformation from the bare Bloch states $\Psi_{\bf k}(n)$ to the 
new eigenstates $\Psi_{\bf k}(n) = \sum_m a_{\bf k}(n,m)
\Phi_{\bf k}(m)$. The system can be diagonalized numerically and 
the CDW order parameters $\chi_{n}=V_{nq_{c}}<\rho_{nq_{c}}>$
are determined self consistently. Once self-consistency is
reached, the eigenvalues $\tilde{E}_{\bf k}(m)$
give the band structure of the system. 

The spectral function in the full BZ then 
consists of a ensemble of weighted delta-functions and is
given by 
\begin{equation} \label{AK}
A_{{\bf k}+n\bf {\bf q}_c}(\omega)=\sum_m a_{\bf k}^2(n,m) 
\delta (\omega-E_{\bf k}(m))
\end{equation}

It should be noticed that this approach mantains its
full validity as long as the initial constraint of no double 
occupancy is satisfied. Therefore we will restrict in the following
to the cases where the CDW order parameter is small enough that
the local density never exceeds one, $\sum_\sigma n_{i,\sigma}\le 1$.

\section{Results}
In this section we will discuss the bandstructure, Fermi surface
and photoemission spectra resulting from either 
1-d or eggbox type CDW scattering. For simplicity, in the following
a planar cell of unitary length ($a=1$) will be used.
Since in the sytems with two CuO$_2$ layers per unit cell
(as in Bi2212)
different stripe orientations in different planes or also
in different regions of the same plane may be realized, we consider
also the case of superimposed (10)- and (01)-stripes.
On the other hand the eggbox-solution, where the symmetry is broken in 
the $x$ and $y$ directions at the same time, can be regarded as a
model for 1-d stripes fastly fluctuating between the (10)- and
(01)-direction, while maintaining a fixed $\vert {\bf q}_c\vert$.
Furthermore, having in mind the possibility of dynamical CDW scattering 
we also investigate
averaged disordered realizations of an eggox charge modulation,
where fluctuations in the absolute length of the scattering vector will
be considered. 
    
\subsection{One-dimensional CDW phase}
The results in this section supplement considerations in
Ref. \cite{SALKOLA} where some effects of 1-d CDW scattering
on the spectral properties were investigated.
Here we analyze the consequences of CDW scattering on the bandstructure 
and, differently from Ref. \cite{SALKOLA}, we start from an
open Fermi surface (induced by 
a next-nearest neighbor hopping term $t'$ as discussed in the
previous section).

\begin{figure}
\hspace{4cm}{{\psfig{figure=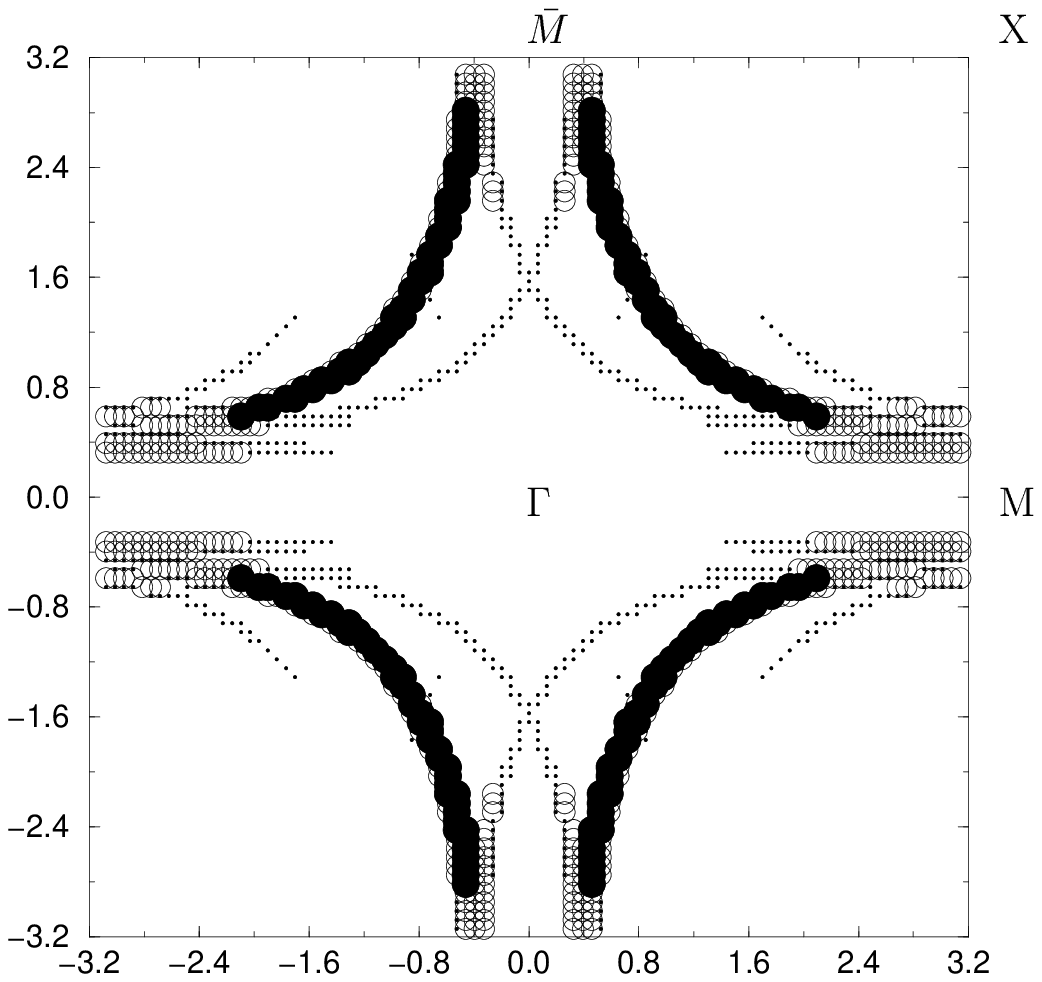,width=4cm,angle=0}}}

\vspace*{3.5cm}

{\small FIG. 1. Fermi surface for a 1-d charge modulation
near the CDW instability where only the first harmonic $\chi_1=0.04$
is different from zero.
Parameters: $\delta=0.2, \alpha=-0.45,
\omega_{0}=0.06eV, g=0.46eV, V_{c}=1.65eV$. CDW modulation $|q_{c}|=0.79$
in direction of $\Gamma - M$.
The plot is for temperature $T=100K$ and the energy window
around E$_{F}$ has choosen to be 50meV. Intensities:
$I>50\%$: full points, $10\%<I<50\%$: circles, $1\%<I<10\%$:
small dots.}
\end{figure}

Fig. 1 displays the calculated Fermi surface (FS) of a one-dimensional 
(1,0)-CDW for a parameter set
close to the instability and doping $\delta=0.2$. Parameters 
such that ${\bf q}_c\approx(\pi/4a,0)$ were chosen, corresponding to
a charge modulation with a period of 8 planar unit cells.
Since we diagonalize the system in a reduced Brillouin zone (BZ) defined
by the choice of $\bf q_{c}$ in eq.\ (\ref{repqc}) the defolding
of the bands leads to a redistribution of spectral weight for the FS states. 
The photoemission intensity for each k-point is given by the following
integration over an energy window around the Fermi energy
$I_k=\int_{E_F-\epsilon}^{E_F+\epsilon} \mbox{d}\omega f(\omega) 
A_{\bf k}(\omega)$ where $f(\omega)$ denotes the Fermi function and
$A_{\bf k}(\omega)$ is the spectral function defined in eq. (\ref{AK}). 
For simplicity the intensity range is divided into three
ranges for high ($I_k>0.5$, full circles), intermediate 
($0.1<I_k<0.5$, open circles) and low ($0.01<I_k<0.1$, dots) intensity.
  
Similar to the results of Ref. \cite{SALKOLA}
there appear displaced shadow bands with low intensity which are
due to q$_c$-scattering processes of the original FS states.
However, since we started with an open Fermi surface (due to
a quasi one-dimensional bandshape around $(\pi,0), (0,\pi)$) the k-space 
regions around the M-points are strongly affected by the scattering.
Moreover, there appears an asymmetry between these two areas since 
around the M-point the scattering is along the FS branches whereas at 
$\bar{M}$ it connects the branches centered around X and Y.
As a consequence more k-states are reduced in intensity around ($\pi,0)$
since the number of states with approximately the same energy
that can be connected by ${\bf q}_c$ is larger than around $(0,\pi)$.
\begin{figure} 
{{\psfig{figure=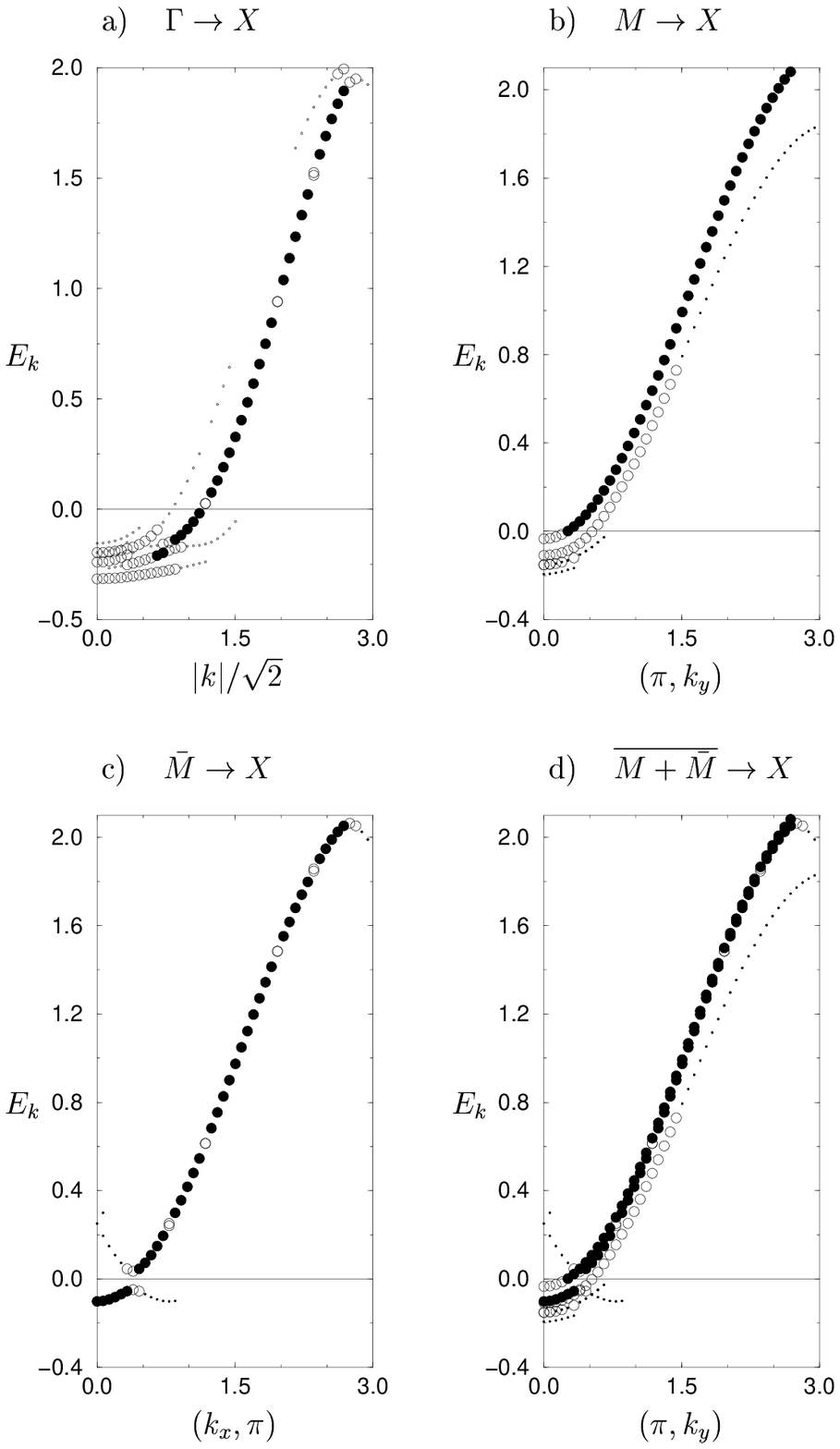,width=7.5cm,angle=0}}}

{\small FIG. 2. (a-c): Cuts of the bandstructure in the full Brillouin zone 
                along various symmetry directions for (10)-scattering, 
                corresponding to the  Fermi surface in
                Fig. 1.  (d): Superposition of the $M\rightarrow X$ and
                the $\bar{M} \rightarrow X$ cut, corresponding to the 
	        Fermi surface in Fig. 1.
                Intensities: $I>50\%$: full points, 
                $10\%<I<50\%$: circles, $1\%<I<10\%$:
                small dots.}
\end{figure}
To analyze the electronic structure in more detail we plot in
Fig. 2 the bands with weight larger than 1\% in the
interesting symmetry directions $\Gamma \rightarrow X$
and $M \rightarrow X$. The intensity range is divided in the
same way as in Fig.\ 1 and the same symbols are used.
Moving along $\Gamma \rightarrow X$ (Fig. 2a) one observes a rich gap structure
around $\Gamma$ which diminishes upon approaching the FS crossing where
besides the main band only the two weak shadow bands survive.
The multiband features around $\Gamma$ are due to the fact that at
the bottom of the band its slope is small and therefore many states with
similar energy can be connected by the CDW vector. 
Taking the cut along $M \rightarrow X$ (Fig. 2b), which is orthogonal
to ${\bf q}_c$, one observes that the CDW has induced the formation of a 
second band which upon approaching X rapidly loses in intensity.
On the other hand the band along the $\bar{M}\rightarrow X$ direction 
(Fig. 2c) displays a CDW gap at the FS crossing since this
direction is parallel to ${\bf q}_c$ and the size of the scattering vector   
is of the same order than the FS branch separation at ($\pi,0$).
\begin{figure}
\hspace*{0.5cm}{{\psfig{figure=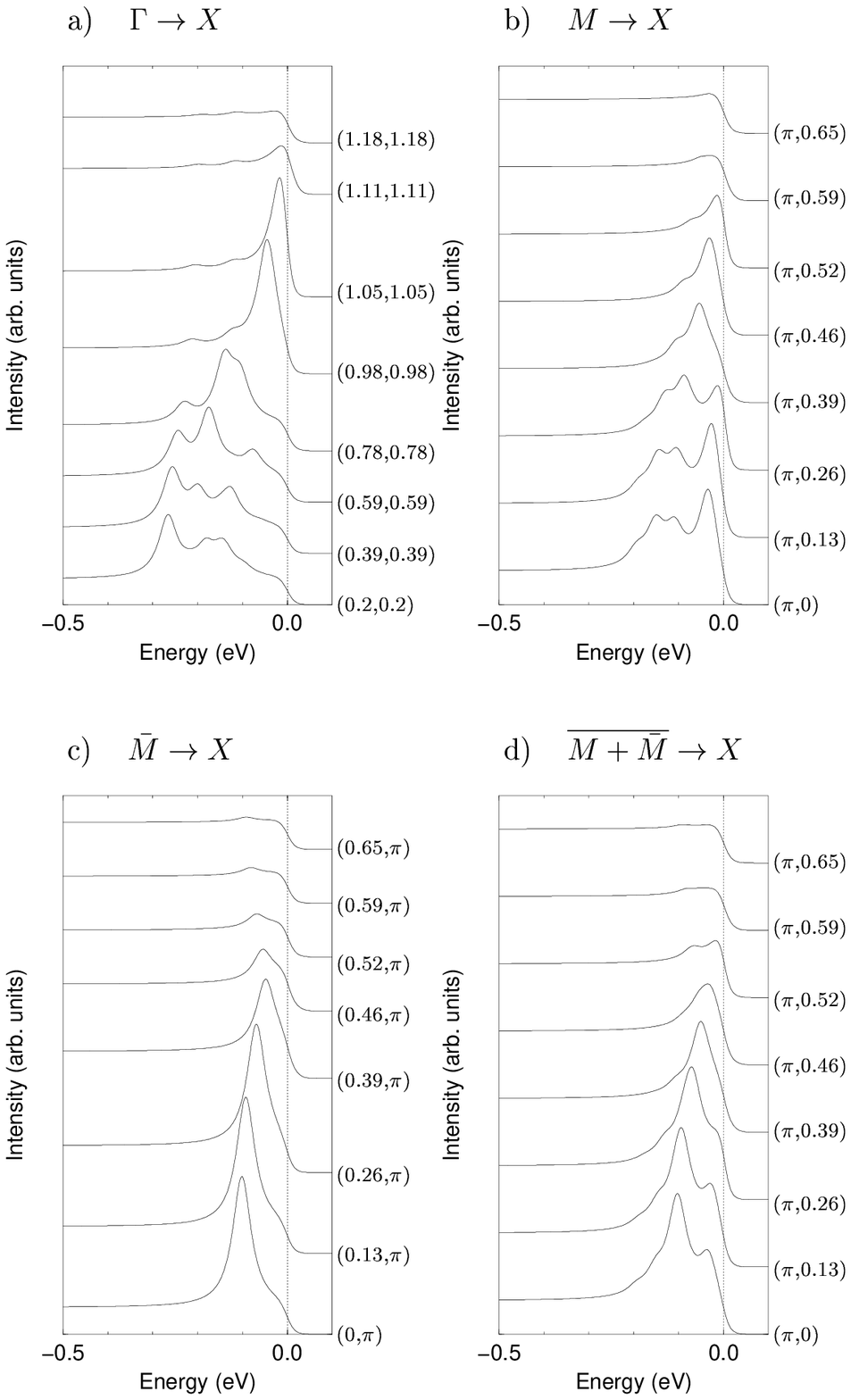,width=7.5cm,angle=0}}}

\vspace*{0.5cm}

{\small FIG. 3. 
                Photoemission spectra corresponding to the cuts in Fig. 3.
                The fermi function has been added as a background.
	        The broadening of the $\delta$-functions is 25meV and the
                temperature T=100K}
\end{figure}
In Fig. 3 we show the energy distribution curves corresponding
to the bandstructure cuts of Fig. 2. Along the $\Gamma \rightarrow X$
direction (Fig. 3a) the gapped structure around $\Gamma$ appears as
a broad multipeaked feature which evolves into a single peak
upon moving towards the X-point.

For the cut along $M \rightarrow X$ one observes two peaks crossing
the FS at $(\pi, 0.39)$ and $(\pi,0.52)$ corresponding to the
two bands in Fig. 2b. The CDW along $\bar{M} \rightarrow X$ displays
in a peak first moving towards the Fermi energy and then bending
back to lower energies (Fig. 3c).
However, this gap no longer can be detected when we consider a system
with superimposed (10)- and (01) CDW scattering. The corresponding
Fermi surface, bandstructure and energy distribution curves are plotted
in Figs. 4,2d,3d respectively. Although there is still reduced weight of the
k-states around $(0,\pi)$, $(\pi,0)$, which are now equivalent,
 the CDW gap at $\bar{M}$ is no
longer visible in Fig. 3d since the averaged bandstructure is dominated
by the states around the M-point.

\begin{figure}
\hspace{4cm}{{\psfig{figure=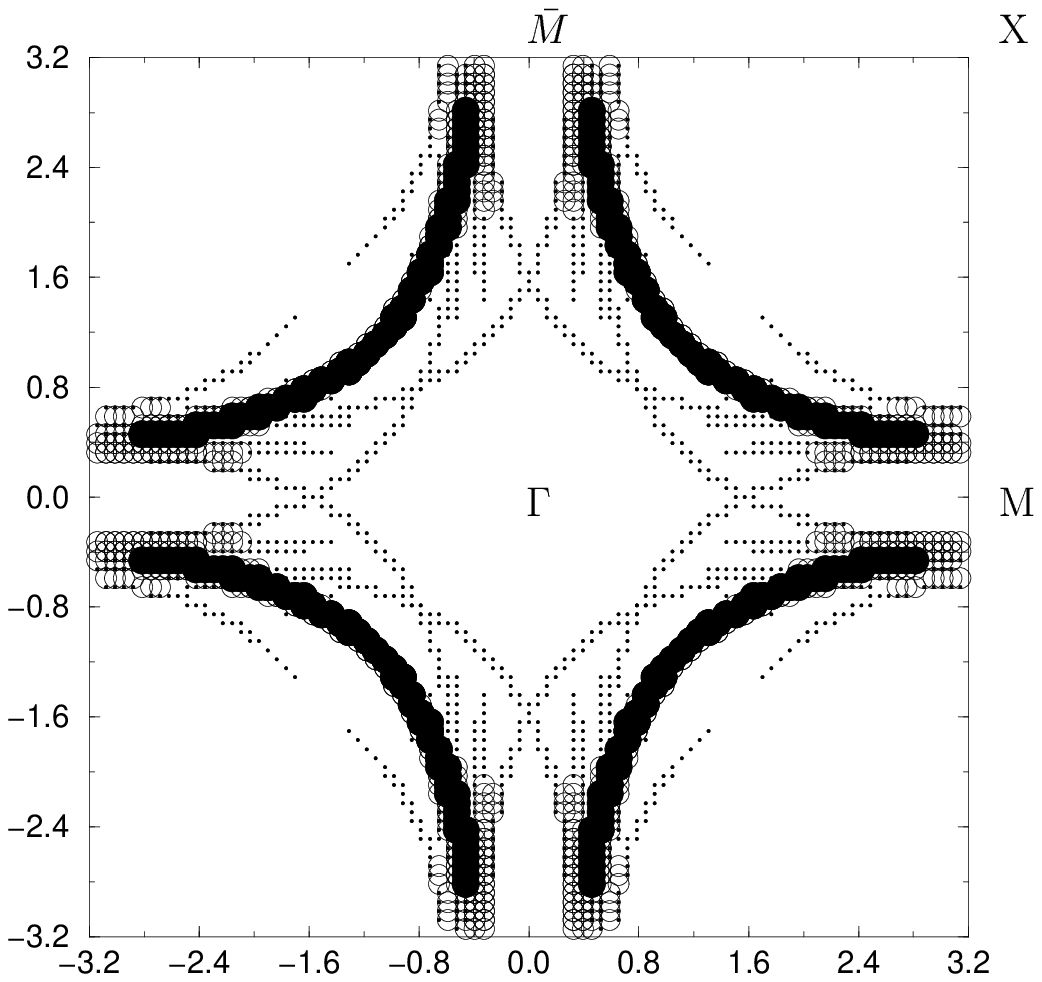,width=4cm,angle=0}}}

\vspace*{3.5cm}

{\small FIG. 4. Fermi surface for the superposition of the (10)-stripe
of Fig. 1 with the equivalent  (01)-charge modulation.
Parameters: $\delta=0.2, \alpha=-0.45,
\omega_{0}=0.06eV, g=0.46eV, V_{c}=1.65eV$. CDW modulation $|q_{c}|=0.79$.
The plot is for temperature $T=100K$ and the energy window
around E$_{F}$ has choosen to be 50meV. Intensities:
$I>50\%$: full points, $10\%<I<50\%$: circles, $1\%<I<10\%$:
small dots.}
\end{figure}

It is also of interest to investigate the band dispersions along the
$\Gamma \to M$ and $\Gamma \to \bar{M}$ directions. Along these
directions a substantial enhancement of the van Hove singularities
have been observed by ARPES experiments \cite{DESSAU,GOFRON,shendessau}
and the obvious question
arises whether this feature can arise from CDW scattering.
Fig. 5 reports in the (a) and (b) sectors the
results of our calculation starting from an unperturbed bandstructure 
(indicated by the diamonds in Fig. 5a,b).
It is apparent that multiple non-dispersive shadow bands
appear along the $\Gamma \to M$ direction as a result of the
CDW scattering, with a substantial redistribution of the 
spectral weight both at higher and lower energies. This latter
occurrence suggests that indeed the broadening of quasiparticle
spectra towards lower energies could experimentally be seen
as a broadening in $k$-space of the area where the band is non-dispersive.
For the presently considered (10) one-dimensional scattering,
the same does not hold in the  $\Gamma \to \bar{M}$ direction,
where the shadow bands are strongly dispersive and much
less numberous (Fig. 5b). 
As a consequence the spectra around $\bar M$ would hardly
look like an enhanced van Hove singularity at low energy.
\begin{figure}
{{\psfig{figure=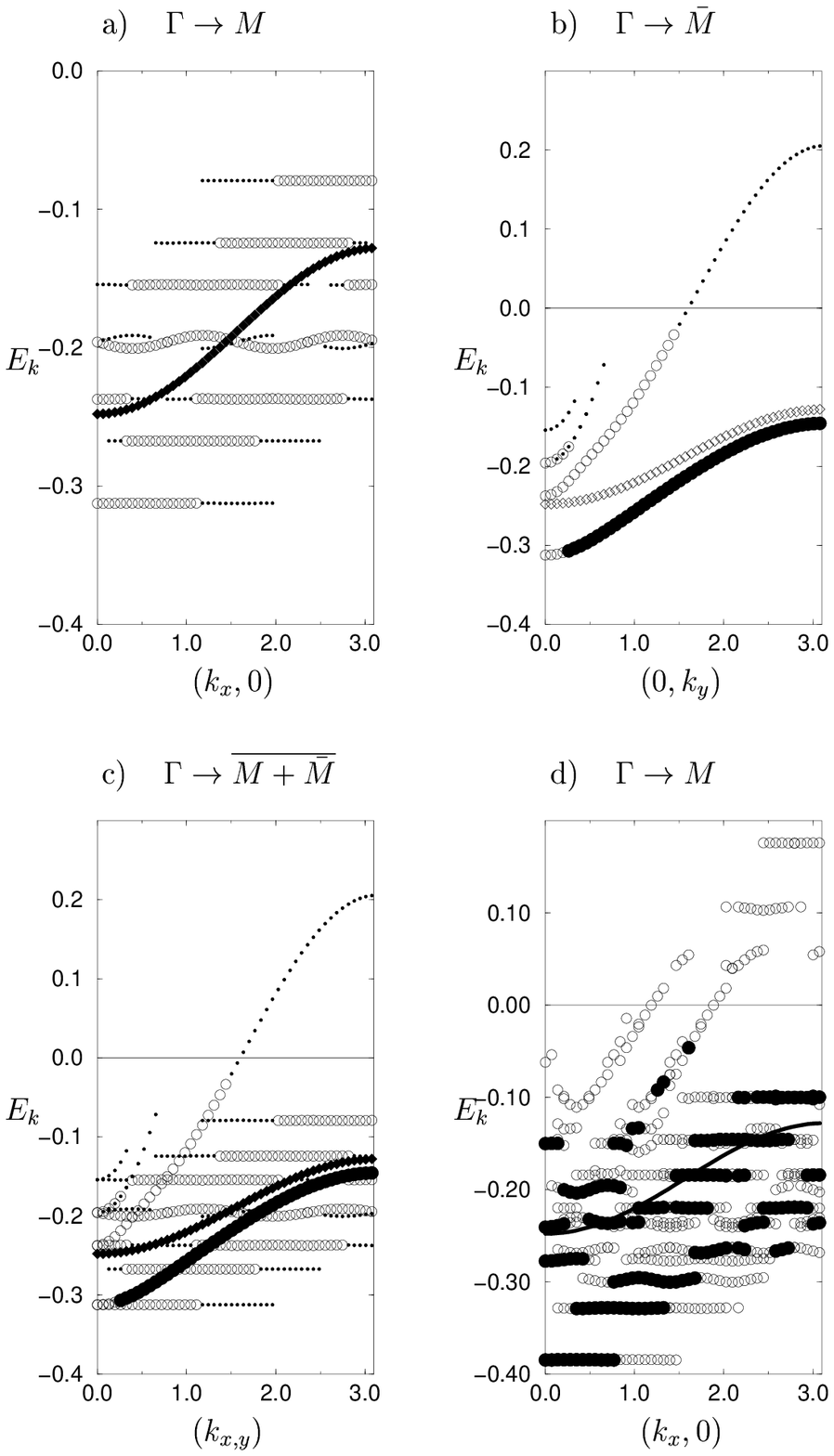,width=7.5cm,angle=0}}}

%\vspace*{1cm}
{\small FIG. 5. (a-c) Bandstructure for one-dimensional CDW scattering
                along the (10)-direction for the same parameters as in 
                Fig. 1 of the paper.
                (a) $\Gamma \rightarrow M=(\pi,0)$, (b) $\Gamma \rightarrow
                \bar{M}=(0,\pi)$, (c) Superposition of (a) and (b). The
                `unperturbed' band is indicated by diamonds. (d) The same
                 cut for 2-d eggbox scattering. Here the `unperturbed' band is
                indicated by the full line.
                Intensities: $I>50\%$: full points, 
                $10\%<I<50\%$: circles, $1\%<I<10\%$:
                small dots.}
\end{figure}
However, the possibility still remains open 
that the spectra of the real materials
can result  from the (nearly static) superpositions
of the spectra in Fig. 5a,b as a consequence of the 
different orientation of the stripes on different $\rm{CuO_2}$ planes or 
on different regions of the same plane. This is depicted in Fig. 
5c. Alternatively, if the
stripes are fastly fluctuating, their effect on the spectra could better
be mimicked by our (static) eggbox solution described below.

\subsection{Ordered eggbox phase}
In this section we are concerned with a two-dimensional charge
modulation resembling the shape of an eggbox.

Fig. 6 displays the calculated FS of a regular 2-d 
CDW for a parameter set
close to the instability and doping $\delta=0.2$.
Similar to Fig. 1 the scattering mostly affects 
the states around $(0,\pi)$ and $(\pi,0)$,
but now of course the features
at M and $\bar{M}$ are symmetric. Moreover, the region of reduced intensity
is much larger than for the superimposed (01)- and (10)-CDW in 
Fig. 4 (note that in both plots the strength of the 
order parameter is the same).
 
\begin{figure}
\hspace{3.8cm}{{\psfig{figure=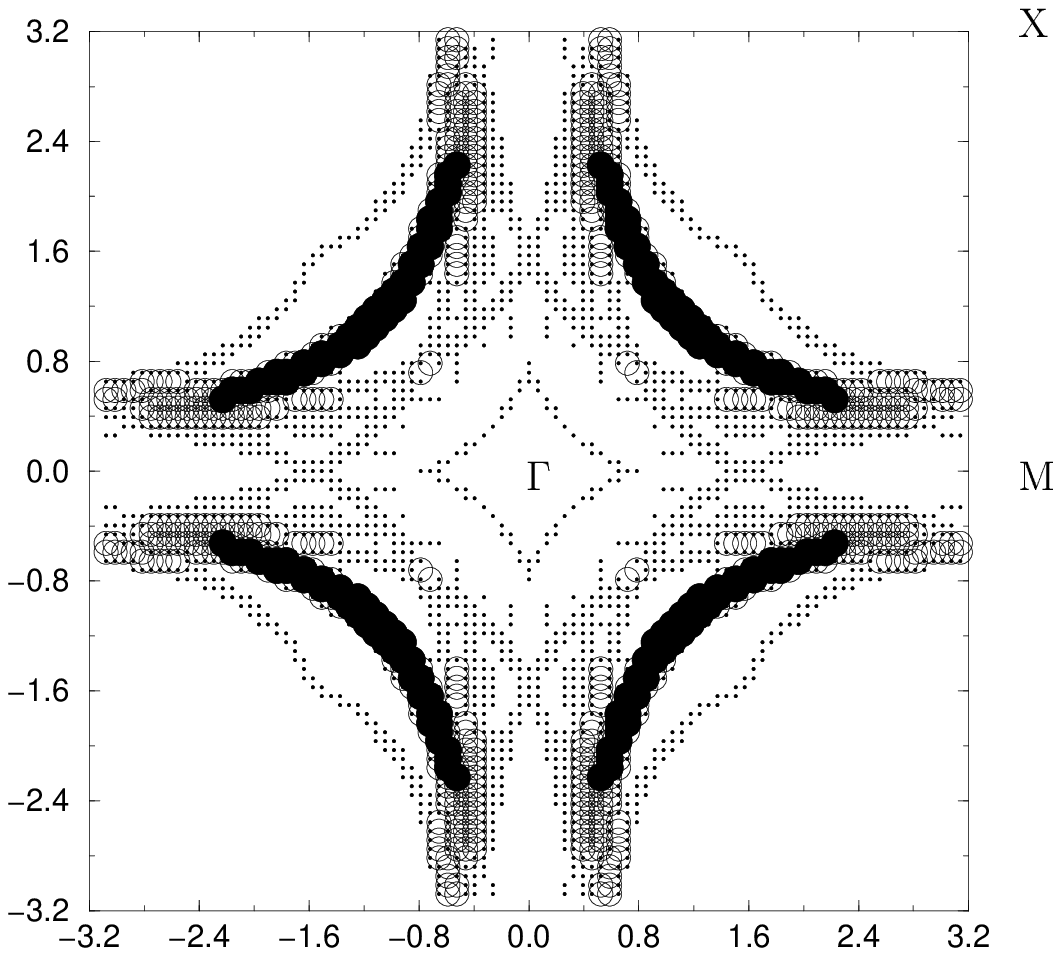,width=4cm,angle=0}}}

\vspace*{3.5cm}

{\small FIG. 6. Fermi surface for an eggbox type charge modulation
near the CDW instability where only the first harmonic $\chi_1=0.04$
is different from zero.
Parameters: $\delta=0.2, \alpha=-0.45,
\omega_{0}=0.06eV, g=0.46eV, V_{c}=1.65eV$. CDW modulation $|q_{c}|=0.79$.
The plot is for temperature $T=100K$ and the energy window
around E$_{F}$ has choosen to be 50meV. Intensities:
$I>50\%$: full points, $10\%<I<50\%$: circles, $1\%<I<10\%$:
small dots.}
\end{figure}
An additional quite interesting feature is the appearance of shadow
Fermi surfaces in the $(1,\pm 1)$ directions, giving rise to ``pockets''.
This feature was already hinted to in the Fermi surface of Fig. 4,
where it was, however less pronounced. 
In particular, here moving from the $\Gamma$ point towards the $X$ (or the
other equivalent points) one first finds some spectral weight at the
Fermi energy for points located around $(0.8,0.8)$ in $k$-space.
Moving further towards $X$ one meets the main Fermi surface with highest 
intensity around $(1.2,1.2)$, and then weak (less than $10\%$ of
intensity) features at $(1.6,1.6)$. It is not unconceivable that
these multiple crossings of the (shadow) Fermi surface(s) have been
observed, although the common interpretation refers to 
magnetic scattering \cite{chubukov}. This finding suggests that, together with
the magnetic scattering, charge scattering could cooperate 
to bring spectral weight at the Fermi surface in a 
``pocket-like'' shape \cite{CAPRARA}.

\begin{figure}
{{\psfig{figure=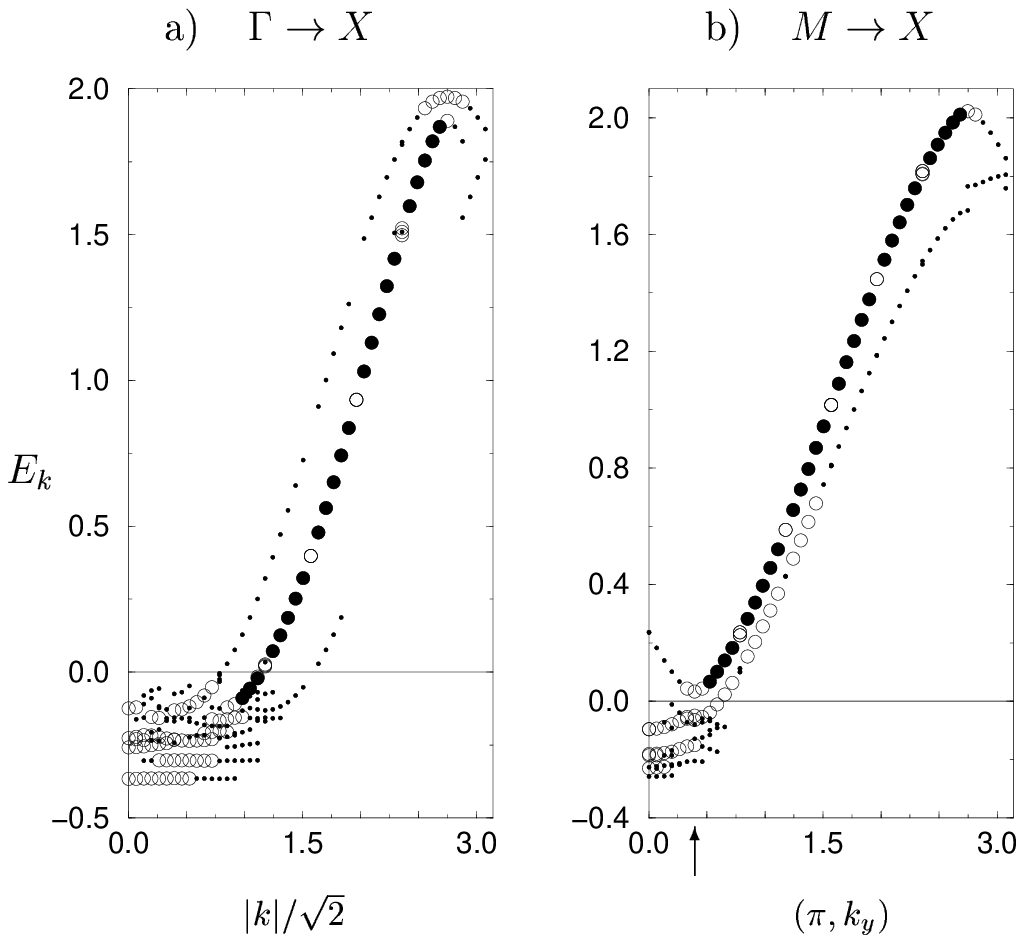,width=7.5cm,angle=0}}}

{\small FIG. 7. Bandstructure in the full Brillouin zone 
                corresponding to the Fermi surface in
                Fig. 1.  Intensities: $I>50\%$: full points, 
                $10\%<I<50\%$: circles, $1\%<I<10\%$:
                small dots.(a) $\Gamma \le k \le X$, 
                (b) $M \le k \le  X$.}
\end{figure}
  
The effect of the 2-d scattering becomes more transparent in the
bandstructure cuts plotted in Fig. 7.
Scanning along the diagonal direction gives a strong
splitting of states around the $\Gamma$ point similar to our previous
findings for 1-d scattering. Moving further towards the X-point 
most of these states rapidly lose in intensity and at the FS crossing
there is only one main band left together with two shadow bands with
low intensity in accordance with the FS shown in Fig. 6.
\begin{figure} 
\hspace{3cm}{{\psfig{figure=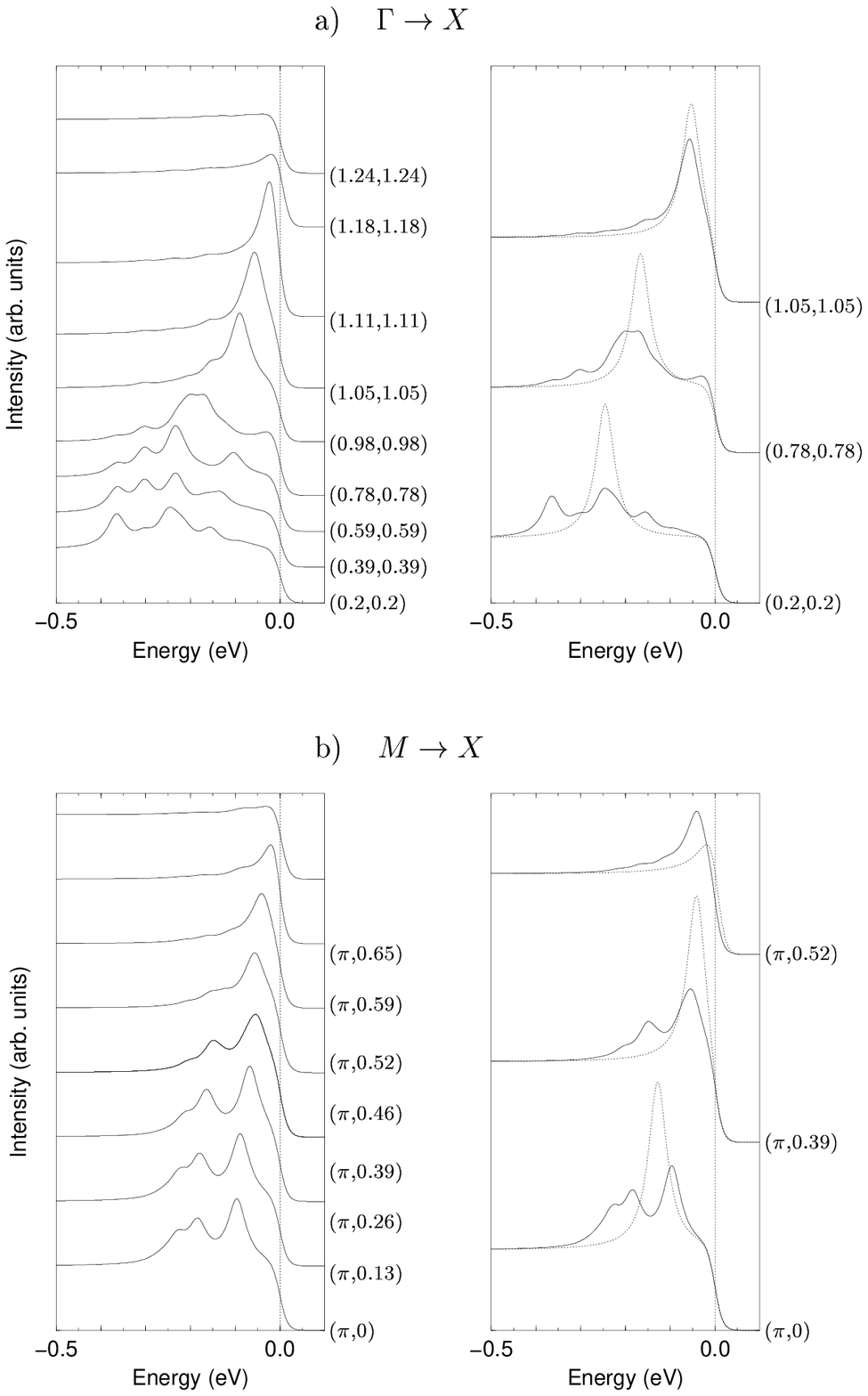,width=5.5cm,angle=0}}}

\vspace{1cm}

{\small FIG. 8. Photoemission spectra for cuts along
               (a) $\Gamma \le k \le X$, 
               (b) $M \le k \le  X$.
                The fermi function has been added as a background.
	        Solid lines: Eggbox solution for the same parameter
                set as in Fig. 1.
		Dashed lines: Homogeneous system.
                The broadening of the $\delta$-functions is 25meV.}
\end{figure}

The band structure along $M \rightarrow X$ is plotted in Fig. 7b.
Also in this case we observe that at the M-point the CDW scattering
has induced various bands with intermediate and low intensity.
At the former FS crossing (indicated by the arrow) now a gap occurs
which is of the order of the CDW order parameter $\chi$.
Although the main band (full circles) bends upwards when approaching
the Fermi level there is a FS crossing by the shadow band with
intermediate intensity. This crossing occurs at a slightly larger 
k$_y$ value than in the unperturbed system which mirrors in Fig. 6
as a broadening of the tube structure around the M-points.

In Fig. 8 we show the energy
distribution curves along the $\Gamma \rightarrow X$
and $M \rightarrow X$ directions. Scanning along the diagonal one
clearly observes the multiple peak structure for the deeply
bound states in Fig. 7a. However, approaching the FS crossing, due
to the rapid loss of intensity in the scattered bands, there appears
a evolution into a single peak structure which crosses the
Fermi level at $k=(1.18,1.18)$. In the right panel of Fig. 8a we
plot a comparison of the CDW spectra with the peak evolution
of the homogeneous system for selected k-points. 
From this plot it becomes clear that along the diagonal direction the peak
structure for the symmetry broken system approaches the unperturbed 
form as one approaches E$_{F}$.
On the other hand, scanning along the M$\rightarrow$X direction, the
peak structure basically is built up by the two bands with intermediate
intensity below E$_F$ which are shown in Fig. 7b.
Therefore one observes a double peak structure at the M-point where
the lower peak decreases in intensity upon approaching the Fermi
level. Due to the induced CDW-gap at the FS crossing of the homogeneous
system $k_y=(\pi,0.52$) there is a shift of the spectra to lower
energy as can be seen from the right panel of Fig (8b). 
It is also quite interesting to see how the CDW-gap 
occurring at k$_y=(\pi,0.52)$ evolves
when one moves along the FS of the unperturbed system.
The corresponding energy distribution curves are shown in Fig. 9.
It turns out that the gap first keeps its value for small angles
up to $\Theta \approx 6^o$. Moving towards the diagonal, the
(pseudo)gap moves at energies below the Fermi level, so that
at the Fermi energy it rapidly decreases and vanishes.
This occurs at $\Theta \approx 20^o$. In addition it is again 
evident that upon approaching the diagonal direction  
the CDW peak structure evolves into the unperturbed one.
These findings may correspond to the experimentally observed shrinking
of the FS \cite{NORMAN}

We now conclude this analysis of the 2-$d$ eggbox CDW scattering
by considering 
the relevant issue of the enhancement of the van Hove singularity
around the $M$ and ${\bar M}$ points. In particular, as it can be seen
in Fig. 5d, one finds a generic broadening of the spectra, 
with dispersionless portions of the spectrum scattered 
at both higher and lower energies with respect to the unperturbed
quasiparticle band. Moreover, two dispersive shadow bands of weak intensity
cross the Fermi energy nearly halfway between $\Gamma$ and $M$.
All these effects contribute to bring spectral weight at 
closer distance from the Fermi level. Once the spectral features
observed in the present simplistic mean-field treatment are
broadened by the fluctuative character of the CDW and by the
many-body effects acting in the real systems, it is not unconceivable that
our present findings correspond 
to an extension of the dispersionless region of the experimental
band structure.

\begin{figure} 
{\hspace{3cm}{\psfig{figure=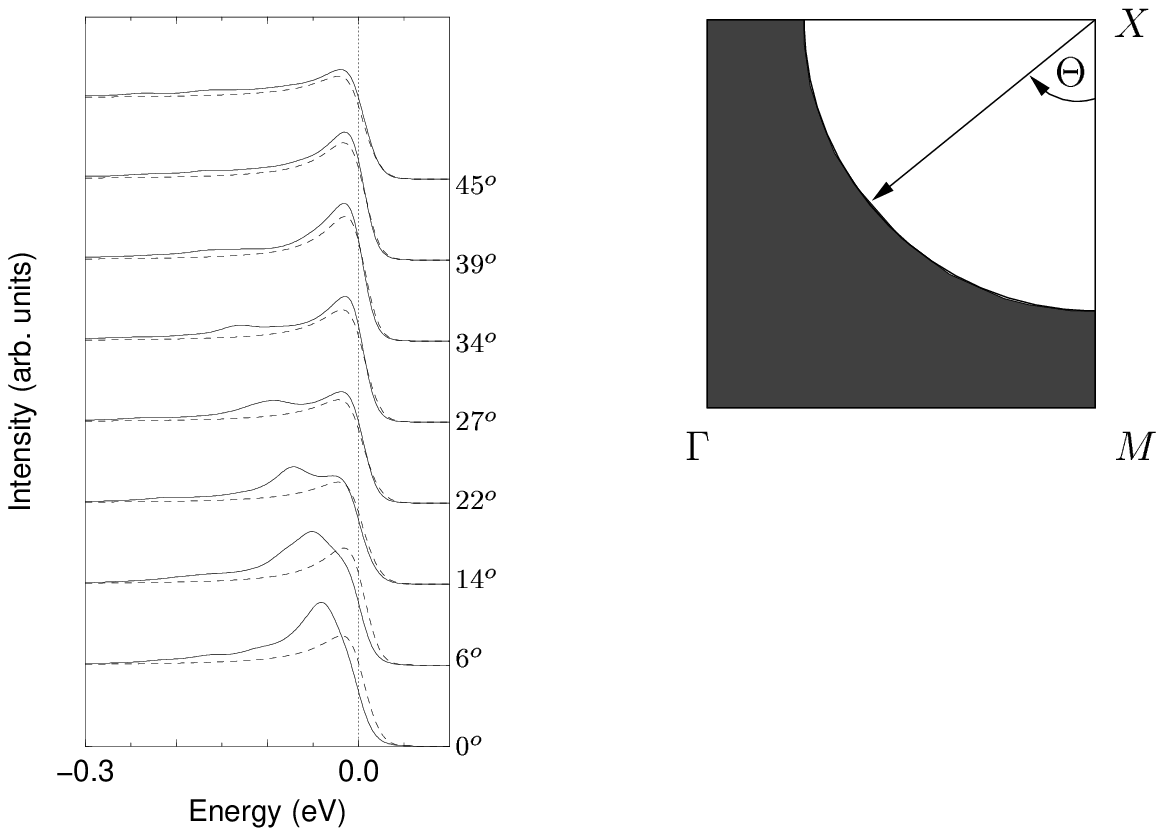,width=5cm,angle=0}}}

\vspace{1cm}

{\small FIG. 9. Scan of the photoemission spectra along the unperturbed
               Fermi surface. The plots are labeled by $\Theta$ which
               is defined in the right panel. Full lines: CDW spectra;
               Dashed lines: Spectra of the unperturbed system.} 
\end{figure}

\subsection{Disordered eggbox phase}
In the preceeding section we have considered an effective one-particle 
Hamiltonian so that all poles of the resulting spectral function
correspond to delta functions. 
We now want to extend our results to have some qualitative idea
of a more dynamical description of the problem,  
i.e. the quasiparticle coupling to collective incommensurate CDW fluctuations. 

As in  Ref.\ \cite{SALKOLA} we start from a phenomenological
one-particle Hamiltonian
\begin{equation}
H=\sum_{k} E_{k}n_{k}+\sum_i V({\bf R_{i}}) n_{i}
\end{equation}
where the kinetic part is the same as in the preceeding section and
$V({\bf R_i})$ represents an effective eggbox potential given by
\begin{equation}
V({\bf R_i})=2 V_0 \sum_n sech(\frac{R_i^x-x_n}{\xi})
                          sech(\frac{R_i^y-y_n}{\xi})
\end{equation}
The amplitude and broadening of the charge modulation is determined by 
$V_0$, $\xi$ and $x_n,y_n$ fix the positions of the individual 
'eggboxes'. In accordance with the ordered eggbox modulation
we choose a mean charge separation of 8a and define 
$x_{n+1}-x_{n}=y_{n+1}-y_{n}=8a+\eta$
where 'a' is the lattice constant and $\eta$ is a random number
varying between $-p a < \eta < p a$. Taking $p=0$ one has
again an ordered eggbox potential whereas upon increasing p (p$_{max}=7$)
long-range charge order is destroyed to an increasing degree.
The charge amplitude $V_0$ is restricted again to values where 
zero double occupancy at each lattice site is preserved. 
The results presented below are averages over
five random configurations calculated by diagonalizing a square lattice
with dimension 40$\times$40 in real space. Doping is again $\delta=0.2$
and the secans-type structure is broadened by $\xi=2$.

Fig. 10 displays the Fermi surface for a disordered eggbox-type
charge modulation. The qualitative features are the same as in Fig. 6,
namely the reduction of intensity around the M-points.
However, the displaced shadow bands are no longer visible now. Instead
the Fermi surface is smeared in k-space i.e. its boundaries consist of
states with low intensity ($< 10\%$). Around the M-points these
states completely fill up the tube structure. This could
correspond to the findings of sequential
angle-scanning photoemission where diffuse features
in these regions have been detected \cite{BIANC3}
and to the suppression of the FS observed in ARPES \cite{NORMAN}.
  
\begin{figure}
\hspace{3.8cm}{{\psfig{figure=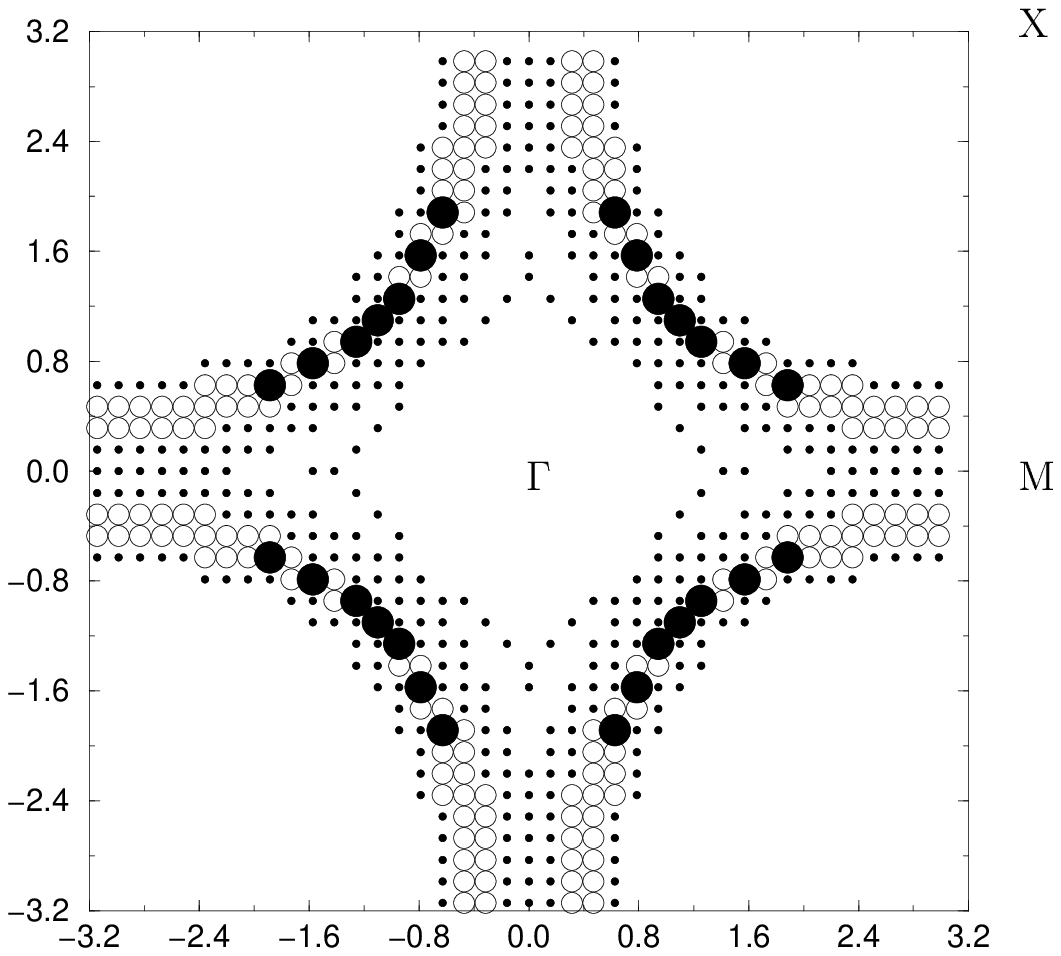,width=4cm,angle=0}}}

\vspace*{3.5cm}

{\small FIG. 10. Fermi surface for a disordered eggbox charge modulation.
Parameters: $\delta=0.2, \alpha=-0.45, p=3,
V_{0}=0.12eV$.
The plot is for temperature $T=100K$ and the energy window
around E$_{F}$ has choosen to be 50meV. Intensities:
$I>50\%$: full points, $10\%<I<50\%$: circles, $1\%<I<10\%$:
small dots.}
\end{figure}
Since we are considering averages over disordered charge configurations 
the energy bands corresponding to Fig. 7 now are also smeared along the 
energy axis therefore removing the fine structure of the CDW gap
topology. This can be seen in Fig. 11 where we have plotted
the energy distribution curves along the $\Gamma \rightarrow X$ and
$M \rightarrow X$ directions in comparison with the homogeneous system
for two different values of the parameter p which defines the
degree of suppression of long-range charge order. This smearing
of the individual CDW peaks is of course more pronounced for $p=7$ in Fig. 
11b than for $p=3$ in Fig. 11a. 
However, scanning along the $\Gamma \rightarrow X$ direction we observe
the same feature than for the ordered eggbox modulation, namely the
evolution of the spectra into a single peak structure upon approaching
the Fermi level thus recovering the FS segment of the quasiparticle. 
On the contrary, at the M-point 
the quasiparticle peak is now completely suppressed and
the spectra are described by a broad feature extending to very low energies.

\begin{figure} 
\hspace{0cm}{{\psfig{figure=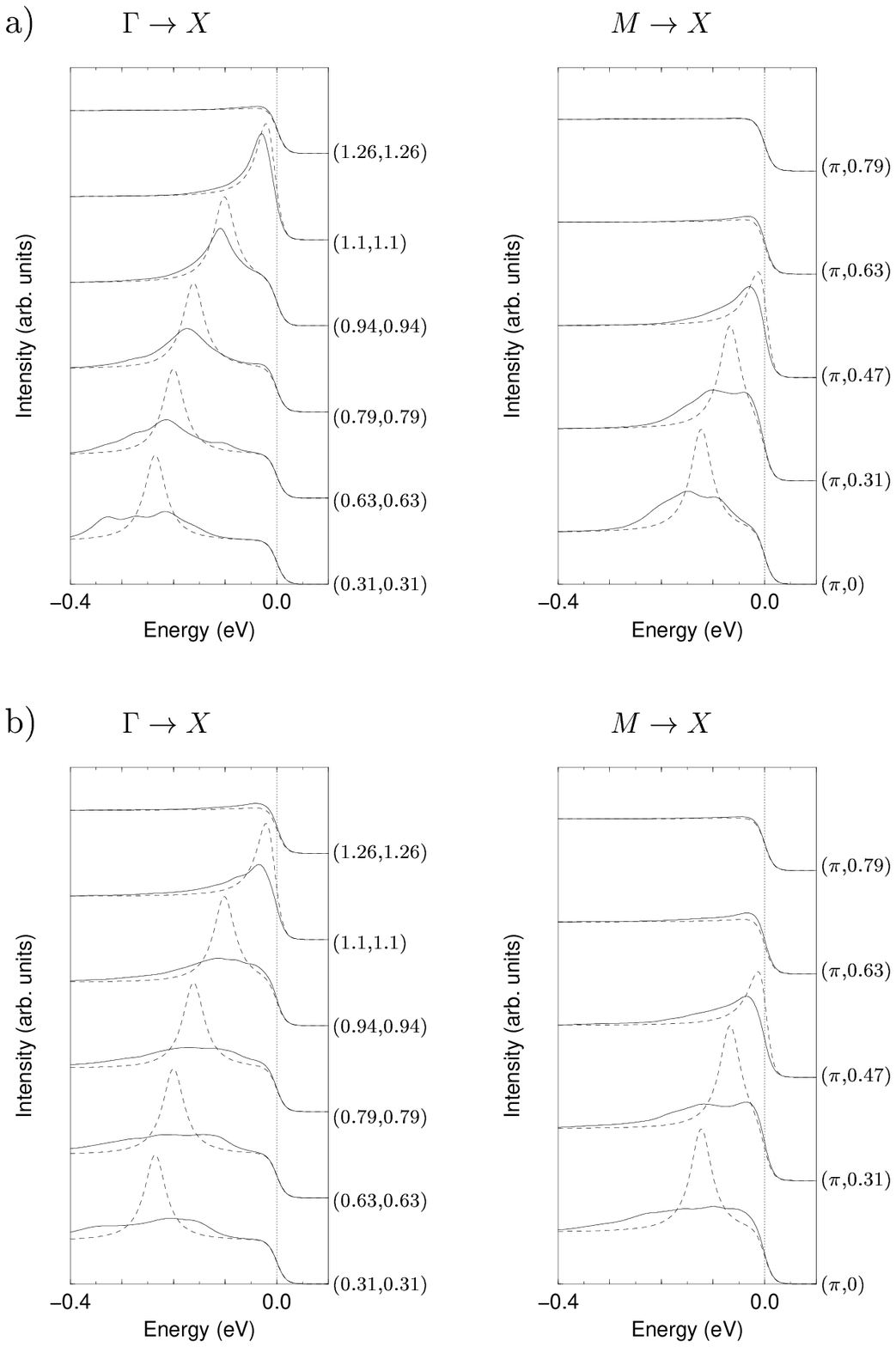,width=8cm,angle=0}}}

\vspace*{0.5cm}

{\small FIG. 11. Photoemission spectra for cuts along
               $\Gamma \le k \le X$ and $M \le k \le  X$ for the
               same parameter set then in Fig. 6.
		The broadening of the $\delta$-functions is 25meV.
	        Dashed lines: Homogeneous system.
		Solid lines: Spectra for the disordered eggbox system.
                The parameter p defing the destruction of long-range 
                charge order is p=3 in (a) and p=7 in (b).}
\end{figure}

\section{Discussion and Summary}
In this paper we have investigated the possible consequences of incommensurate
CDW scattering with regard to the unusual Fermi surface and photoemission 
features in underdoped bismuth cuprates already mentioned above. Our  results 
have been obtained within a scheme, which is subjected to some limitations
both at the level of the starting model and of its treatment.
In particular,  only the physics of CDW
modulations has been  considered, thus neglecting the relevant
interplay between charge, spin, and Cooper pair fluctuations.
Nevertheless it is quite interesting and instructive by itself
that our treatment of a pure CDW symmetry broken system still
captures some features of the observed spectral properties.
This indicates that charge-ordering can indeed play a role
in determining the properties of the cuprates.

The simplified model considered here 
has been approximately treated in a 
mean-field scheme aiming to capture 
features of well-formed locally ordered stripes in the underdoped cuprates.
In this scheme the fermionic quasiparticles do not
interact neither among themselves nor with the collective
CDW fluctuations. Instead the static symmetry breaking due to 
charge modulation produces Bragg scattering giving rise to
multiple bands and gap opening. As a result the bands
arising from the mean-field description are due to $\delta$-like
quasiparticle peaks, which obviously do not individually correspond to 
the real experimental features. Nevertheless, it is worth noting that
the multiple Bragg scattering due to incommensuration leads for most $k$-points
to the appearance of multiple quasiparticle peaks [the shadow bands,
see Figs. (2),(5),(7)].
It is quite natural that the broadening arising from the 
scattering between quasiparticles and the scattering between 
quasiparticles and collective modes 
will mix these peaks thus producing
the  broad features commonly detected in photoemission experiments.
Still we believe that our analysis provides
the location and relative position of the spectral features.

We considered one-dimensional charge modulations along the (1,0) and/or
(0,1) as well as a two-dimensional eggbox-like texture. 
The one-dimensional solutions could account
for some enhancement of the van Hove singularities, but
do not seem particularly successful in describing the
appearance of a pseudogap near the $M$ and $\bar M$ points.
On the other hand the eggbox case turns out  to be particularly appealing. 
The main findings are that the 2-d eggbox CDW scattering might
account both for a flattening of the band dispersion in the $\Gamma \to M$
and  $\Gamma \to \bar M$ directions and for the arising of a leading-edge
gap around the $M,\bar M$ points leaving finite portions of
the Fermi surface gapless. As discussed in the
introduction, this latter quite unusual and non BCS-like behavior of the
gap has indeed been observed \cite{NORMAN} and suggests that 
the charge fluctuations in the particle-hole channel could substantially
participate to the spectral features around the $M,\bar M$ points.
In this regards our finding supports the idea that
 the scattering induced by a charge ordered superstructure
works in many aspects in the same direction than mechanisms involving
incoherent pairing correlations (possibly cooperating with
magnetic scattering \cite{CAPRARA}).
On the other hand, we also found  that a disordering of the
eggbox structures ``washes out'' the weak low-energy features and
brings about the disappearance of the
leading-edge gap due to charge scattering, but preserves the 
feature of Fermi surface only formed by disconnected arcs.
In this last situation, the pseudogap will only appear as a
(robust) suppression of the spectral weight, which is shifted
at higher binding energies. 

The overall success of the eggbox CDW scattering in reproducing some
highly non-trivial features of the single-particle spectra of the
underdoped cuprates suggests that a dynamical charge pseudo-ordering
is a relevant aspect of the physics of these materials.

\acknowledgments
G.S. acknowledges financial support from the Deutsche Forschungsgemeinschaft
as well as  hospitality and support from the Dipartimento
di Fisica of Universit\`a di Roma ``La Sapienza''where part of this work
was carried out. This work was partially supported by INFM-PRA (1996).

\end{multicols}


\begin{thebibliography}{99}
\bibitem{notaps}The reason why phase separation nevertheless is 
	experimentally observed in
	the La$_{2}$CuO$_{4+y}$ oxides is that the negatively 
	charged oxygen ions are sufficiently mobile in these materials, 
	thus screening the
	positive holes and compensating the charge unbalance.
\bibitem{notainstab} The occurence of the CDW instability
	in the present effective model is in good qualitative
	and quantitative agreement with the instability found in
	\cite{CAST,BECCA}, where the scattering amplitude of the
	model eq.(\ref{HUHOL}) has been calculated systematically up to
	order 1/N in the spin degeneracy N.
\end{thebibliography}
\end{document}